\documentclass{article}
\usepackage{xcolor}
\usepackage{spconf}
\dimen\footins=8\baselineskip\relax
\usepackage[bottom]{footmisc}
\usepackage{graphicx}
\usepackage{amssymb,amsmath,bm}
\usepackage{textcomp}
\usepackage{graphicx}
\usepackage{wrapfig}
\usepackage{lipsum}
\usepackage{dblfloatfix}
\usepackage{fixltx2e}
\usepackage{nameref}
\usepackage{hyperref}
\usepackage{graphicx}
\usepackage{algorithm}
\usepackage{varioref}
\usepackage{booktabs}  % nice looking tables
\usepackage{siunitx}
\usepackage[noend]{algpseudocode}
\usepackage{numprint}
\usepackage{textcomp}
\usepackage[T1]{fontenc} 
\usepackage[noend]{algpseudocode}
\usepackage{tablefootnote}

\makeatletter
\def\BState{\State\hskip-\ALG@thistlm}
\makeatother
\usepackage{xpatch}
\xpatchcmd{\algorithmic}
{\ALG@tlm\z@}{\leftmargin\z@\ALG@tlm\z@}
{}{}
\title{Robust feature clustering for unsupervised speech activity detection}
\name{Harishchandra Dubey, Abhijeet Sangwan, John H. L. Hansen\textsuperscript{+}\thanks{\textsuperscript{+}This project was funded in part by AFRL under contract FA8750-15-1-0205 and partially by the University of Texas at Dallas from the Distinguished University Chair in Telecommunications Engineering held by J. H. L. Hansen.}\thanks{\textcolor{blue}{This material is presented to ensure timely dissemination of scholarly and technical work. Copyright and all rights therein are retained by the authors or by the respective copyright holders. The original citation of this paper is:
H. Dubey, A. Sangwan, and J. H. L. Hansen, "Robust feature clustering
for unsupervised speech activity detection,"in IEEE ICASSP, 2018, pp.
2726-2730.}}}
\address{Robust Speech Technologies Lab, Center for Robust Speech Systems\\
The University of Texas at Dallas, Richardson, TX 75080, USA \\
	{\small \tt \{Harishchandra.Dubey, Abhijeet.Sangwan, John.Hansen\}@utdallas.edu}}

\begin{document}
\maketitle
\vspace{-2mm}
\begin{abstract}
\vspace{-2mm}
In certain applications such as zero-resource speech processing or very-low resource speech-language systems, it might not be feasible to collect speech activity detection (SAD) annotations. However, the state-of-the-art supervised SAD techniques based on neural networks or other machine learning methods require annotated training data matched to the target domain. This paper establish a clustering approach for fully unsupervised SAD useful for cases where SAD annotations are not available. The proposed approach leverages Hartigan dip test in a recursive strategy for segmenting the feature space into prominent modes. Statistical dip is invariant to distortions that lends robustness to the proposed method. We evaluate the method on NIST OpenSAD 2015 and NIST OpenSAT 2017 public safety communications data. The results showed the superiority of proposed approach over the two-component GMM baseline. 
\end{abstract}
%NIST OpenSAT 2017 public safety communications (PSC) task and peer-led team learning etc,
%Proposed technique is deterministic and do not require parameter optimization.
%
%Unsupervised SAD for real-world data corrupted with high levels of noise and other distortions is a challenging task. Among unsupervised/semi-supervised SAD methods, thresholding or fitting a Gaussian mixture model in feature space is a common practice~\cite{sholokhov2018semi}. Further, under the influence of naturalistic distortions, the features may not be accurately modeled by a Gaussian distribution.
%
%Several interesting studies exist on the topic, however there are still significant challenges.
%The results and discussions shows that dip is an attractive tools for clustering speech features.
%
%In this study, we introduce and evaluate a method for unsupervised SAD based on Hartigan's dip test. The speech/non-speech decision is based on dithree language andp-statistics derived from empirical cumulative distribution of the speech features. The proposed approach is evaluated on adversely degraded acoustic channels of NIST OpenSAD 2015 data. 
%  
%The proposed approach for SAD is SAD a is based on statistical dip derived from empirical cumulative distribution function of speech features. We used the combosad, long-term spectal divergence and energy based features for evaluating the proposed SAD  approach.  
%
\vspace{-2mm}
\section{Introduction}
\vspace{-2mm}
Speech activity detection (SAD) is an essential front-end in most speech systems such as automatic speech recognition, speaker verification~\emph{etc}~\cite{sholokhov2018semi}. SAD methods are broadly considered into two categories: (1) supervised and (2) unsupervised. While supervised approaches are trained on massive amount of annotated data, unsupervised techniques do not require labeled data~\cite{sohn1999statistical}. Supervised techniques tend to perform poorly on mis-matched train and test conditions. Gaussian mixture models (GMMs) have been extensively used for supervised, semi-supervised and unsupervised SAD~\cite{sholokhov2018semi,sadjadi2013unsupervised,dubey2016robust}. 
%The proposed approach is both non-parametric and does not make modal assumptions such as in any GMM approach. 
%It is an iterative method based on Hartigan dip test recursions. We compare the proposed SAD technique with a two-component GMM. Statistical dip based SAD makes no assumption on the distribution of speech features. The approach takes speech features and performs an iterative search for maxima in  the statistical dips of the feature distribution. 
%he approach can be used with a variety of speech features, thus complementing the discriminative features with dip-based decision.
%
%\vspace{-7mm}
%\section{Background}
%\vspace{-4mm}
%\vspace{-2mm}
Robust SAD over degraded channels have been of interest for several years~\cite{ramirez2004efficient,ramirez2005statistical,zhang2013deep,ghosh2011robust, shin2010voice,gorriz2006hard,dubey2017using}. SAD methods are varied, from energy-based~\cite{sohn1999statistical} to deep neural networks (DNN)~\cite{zhang2013deep}. The DARPA RATS program supported the SAD research in multiple phases that led to the development of advanced approaches~\cite{ng2012developing,saon2013ibm,thomas2015improvements,graciarena2013all,novotney2016bbn,graciarena2016sri}. Recent work in~\cite{sholokhov2018semi} summarized the SAD developments in context of semi-supervised and unsupervised techniques. Specifically, it introduced the idea of semi-supervised learning in conventional expectation-maximization (EM) algorithm for semi-supervised GMM for speech activity detection. 
\vspace{-2mm}
\section{Proposed Method}
\vspace{-2mm}
\subsection{Feature Extraction}
\vspace{-2mm}
The handcrafted five-dimensional features for Combo-SAD approach were introduced in~\cite{sadjadi2013unsupervised}. Authors performed mean and variance normalization on each feature dimension. The normalized features were later processed with principal component analysis (PCA) for extracting the first principal component that was named \textit{Combo} feature. The Combo features were later employed to consider a two-component GMM for unsupervised SAD~\cite{sadjadi2013unsupervised}. We used the two-component GMM as baseline decision backend for comparison with proposed Dip-based unsupervised backend in this study.
%%
%
%\begin{minipage}{.5\linewidth}
\setlength{\textfloatsep}{10pt}% Remove \textfloatsep
\noindent
\begin{algorithm}[!t]
	\caption{\strut~\textit{computeDip}}\label{computedip}
	\textbf{Input:} speech features were sorted in ascending order i.e., \textbf{o}=[$o_{1}$, $o_{2}$, ...,$o_{N}$] where $o_{1} \le o_{2} \le ... \le o_{N} $. \\
	\textbf{Output:} primary modal interval [$o_{L}$, $o_{U}$], DIP and p-value, $p$. \\
	%	\hspace*{\algorithmicindent} \textbf{Input:} speech features sorted in ascending order i.e., \textbf{o}=[$o_{1}$, $o_{2}$, ...,$o_{N}$] where $o_{1} \le o_{2} \le ... \le o_{N} $. \\
	%	\hspace*{\algorithmicindent} \textbf{Output:} primary modal interval [$o_{L}$, $o_{U}$], DIP and p-value, $p$. \\
	\begin{algorithmic}[1]
		%\Function{Dip Test}{}
		\vspace{-2mm}
		%\lipsum[4-6]
		\Statex \textbf{Step 1:} Initialize, lower point $o_{L}$= $o_{1}$, upper point $o_{U}$= $o_{N}$ and D = 0.
		\Statex \textbf{Step 2:} Compute  greatest convex minorant $G$ and least concave majorant $H$ of empirical distribution $F$ of features in interval [$o_{L}$, $o_{U}$]~\cite{HarP85}. Let the points of contact with $F$ are respectively, $g_{1}$, $g_{2},..,g_{k}$ (for $G$) and $h_{1}$, $h_{2},..,h_{m}$ (for $H$). 
		\Statex \textbf{Step 3:} Let $d$= max |$G(g_{i})$ - $H(g_{i})$| > max |$G(h_{j})$ - $H(h_{j})$| and the maximum occurs at $h_{j} \le g_{i} \le h_{j+1}$. Then, define $o_{L}^{0}$= $g_{i}$, $o_{U}^{0}$= $h_{j+1}$. 
		\Statex \textbf{Step 4:} Let $d$= max |$G(h_{j})$ - $H(h_{j})$| $\ge$ max |$G(g_{i})$ - $H(g_{i})$| and the maximum occurs at $g_{i} \le h_{j} \le g_{i+1}$. Then, define $o_{L}^{0}$= $g_{i}$, $o_{U}^{0}$= $h_{j}$. 
		\Statex \textbf{Step 5:} If $d \le D$, Stop and set DIP= $\frac{D}{2}$.	
		\Statex \textbf{Step 6:} If $d > D$, set D= \text{max} \{ $$\sup_{o_{L} \le o \le o_{L}^{0}} | G(o) - F(o)|, 
			\sup_{o_{U} \le o \le o_{U}^{0}} | H(o) - F(o) |$$ \}, where $\sup$ is the supremum (supremum is the smallest number that is greater than or equal to every number in the set).	
		\Statex \textbf{Step 7:} Set $o_{L}$ = $o_{L}^{0}$,  $o_{U}$ = $o_{U}^{0}$. Go to Step 2. 
	% \EndProcedure
	% \footnotetext[17]{This is my footnote!}
	% \EndFunction
	\end{algorithmic}
\end{algorithm}		
\vspace{-3mm}
\subsection{Hartigan dip test}
\vspace{-1mm}
%
%The dip test defines "unimodality" as:~\textit{a feature distribution is unimodal if its cumulative distribution is of convex type upto its modal interval and concave type after the modal interval}. 
%
The dip test~\cite{hartigan1985dip} is a statistical test for hypothesizing the modality of a distribution. It is based on the geometrical shape of the feature distribution. The dip test tries to fit a piecewise linear function, that is convex then a concave, to the cumulative distribution. The unimodality is decided based on the~\textit{goodness of this piecewise linear fit}~\cite{HarP85}. We leveraged recursions based on dip test for clustering feature space into speech and non-speech classes. This paper is motivated by the recent success in applying Hartigan test for clustering extremely noisy data from other domains~\cite{maurus2016skinny}. Application to speech processing, particularly speech activity detection is a novel contribution of this paper. By comparing the dip statistics with that of a suitable reference unimodal distribution (i.e., null distribution), a p-value is set for the null hypothesis. Using the significance level, $\alpha= 0.05$, we may reject or favor the null hypothesis (unimodality) against the alternative hypothesis (multi-modality). In this way, the dip test quantifies the empirical cumulative distribution's departure from unimodality. Importantly, the dip test (see Algorithm~\ref{computedip}~\textit{computeDip}) communicates the modal interval [$o_{L}$,$o_{U}$], the p-value and the DIP. 
%This interval refers to the region where the empirical cumulative distribution has the maximum slope~\cite{maurus2016skinny}. 
%s. The dip  statistics can help in accepting or rejecting the hypothesis that the feature distribution is unimodal. It does not directly count the number of clusters and therefore not a direct estimation of locations of speech/non-speech boundaries. 
%
%We now present the algorithmic details of the dip test.
%
It is important to note that the proposed clustering approach works on all frames of a single utterance thus it a utterance-level approach. The speech feature vector, $\mathbf{feats}$ are sorted in increasing order. We still store the original feature vector in memory for preserving the temporal order (time information) of the frames. Let the sorted features (observations) be $\mathbf{o}= o_{1}$, $o_{2}$,...,$o_{N}$ with $o_{1}$ $\le$ $o_{2}$ $\le $...$\le$ $o_{N}$ where $N$ is the length of the feature vector (number of frames). All speech and non-speech modal intervals, ($o_{i}$, $o_{j}$) in the feature space would be the pairs of values from $\mathbf{o}$. If $N$ is the length of $\mathbf{feats}$ or equivalently $\mathbf{o}$, total number of possible modal intervals would be $\binom{N}{2}$ = $\frac{N(N-1)}{2}$ that is combinations obtained by choosing two values out of $\mathbf{o}$ vector. Now, for each modal interval ($o_{i}$, $o_{j}$) we compute the greatest convex minorant, $G$ of empirical distribution, $F$ in (-$\infty$, $o_{i}$) and least concave majorant, $H$  of empirical distribution, $F$ in ($o_{j}$,$\infty$). Let $d_{ij}$ be the maximum distance between $F$ and curves $G$, $H$ in modal interval ($o_{i}$, $o_{j}$). Then, the DIP is given as 
\begin{equation}
\text{DIP} = \frac{1}{2} \text{min}  \{ d_{ij} \} ,
\label{eq1}
\end{equation}
over all modal interval ($o_{i}$,$o_{j}$) such that the line segment from [$o_{i}, F(o_{i})$ + $\frac{1}{2}d_{ij}$] to [$o_{j}$, $F(o_{j}) - \frac{1}{2}d_{ij}$] lies in the set defined by
\begin{equation}
\{ o, y    | o_{i}  \le o \le o_{j}, F(o) - \frac{1}{2} d_{ij} \le y \le F(o) + \frac{1}{2} d_{ij}   \}
\label{eq2}
\end{equation}
The Equation~\ref{eq2} ensures that the greatest convex minorant, modal segment and the least concave majorant together form a unimodal distribution. The Algorithm~\ref{computedip}~\textit{computeDip} compute the DIP value, the modal interval and the p-value, $p$ from the significance test. 
%This routine has a time complexity of $\mathcal{O}(N)$ for a feature vector of length $N$.
%%
\setlength{\textfloatsep}{0pt}% Remove \textfloatsep
\begin{figure}[!t]
\centering
\includegraphics[width=210pt]{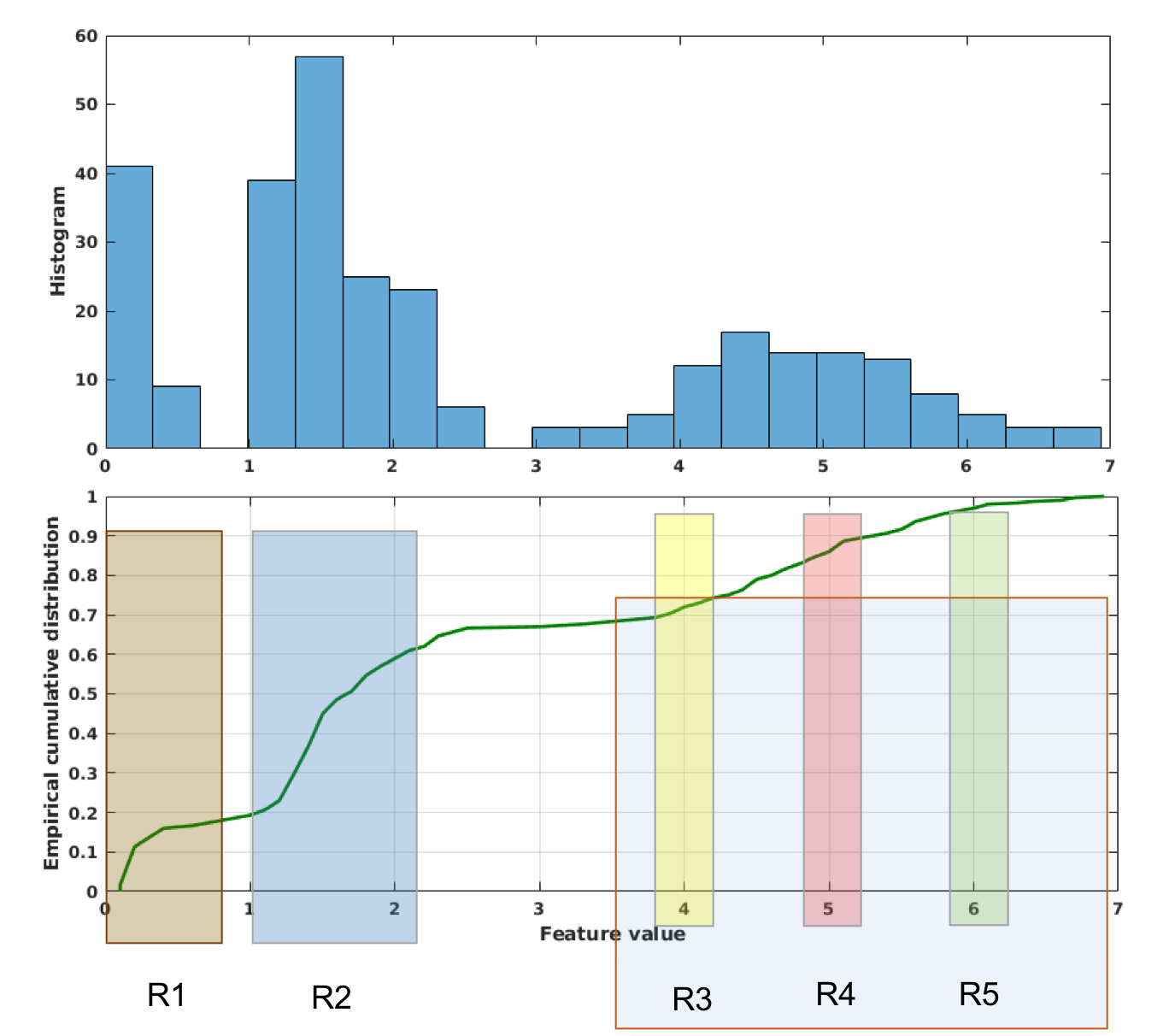}
%\vspace{-2mm}
\caption{Illustration of the dip-based clustering technique on synthetic data with five classes, identified with R1 to R5 where three regions R3, R4 and R5 lie close to each other in the feature space.}
\label{fig_illus}
%\vspace{-5mm}
\end{figure}
%
%\lipsum[3-4]
\setlength{\textfloatsep}{0pt}% Remove \textfloatsep
	\noindent
\begin{algorithm}[!t]
\caption{\textit{Dip-SAD}}\label{dipsad}
%\hspace*{\algorithmicindent} \textbf{Input:} speech features for SAD \\
%\hspace*{\algorithmicindent} \textbf{Output:} speech non-speech segmentation \\
\textbf{Input:} frame-level speech features from an utterance \\
\textbf{Output:} speech non-speech labels for each frame\\
\begin{algorithmic}[1]
\vspace{-2mm}
	\floatname{algorithm}{Procedure}
	\renewcommand{\algorithmicrequire}{\textbf{Input:}}
	\renewcommand{\algorithmicensure}{\textbf{Output:}}
	%\Procedure{Dip-based Segmentation}{}
		\Statex \textbf{Step 1:} Sort the features in ascending order and let \textbf{o}=[$o_{1}$, $o_{2}$, ...,$o_{N}$] be the ordered vector, where $o_{1} \le o_{2} \le ... \le o_{N} $. The significance level, $\alpha$ is set to 0.05 for all experiments reported in this paper. 
		\Statex \textbf{Step 2:} $\textit{\{$o_{L}$, $o_{U}$, $p$\}} \gets \text{\textit{computeDip}(\textbf{o}) }$
		\Statex \textbf{Step 3:} If $p$ > $\alpha$, then the detected primary modal interval is [$o_{L}$, $o_{U}$]. Else, [$o_{1}$, $o_{N}$] is primary modal interval.
		\Statex \textbf{Step 4:} Recurse into the modal interval to find the list $I_{mid}$ of the modal intervals within detected primary mode. 
		\Statex \textbf{Step 5:} Now, we check to the right and left of the primary modal interval recursively and extract additional modes if found. 
		%\skip\footins=-\bigskipamount % or something else
		
		%At this stage, we have at-least three possible segments, [$o_{1}$,$o_{L}$], [$o_{L}$,$o_{U}$], and  that have to be rechecked for modality. 
		%	
		\Statex \textbf{Step 6:} $\textit{\{$u$\}} \gets \underset{o_{U}\in I_{mid}} {\mathrm{min}} \text{ (\textbf{$o_{U}$}) }$, $\textit{\{$l$\}} \gets \underset{o_{L}\in I_{mid}} {\mathrm{max}} \text{ (\textbf{$o_{L}$}) }$. 
		\Statex \textbf{Step 7:} $\textit{$p_{l}$} \gets \mathrm{\textit{computeDip}}\text{(\textbf{ $\forall o_{j}: o_{j} \le u$}) }$ , $\textit{$p_{u}$} \gets \mathrm{\textit{computeDip}}\text{(\textbf{ $\forall o_{j}: o_{j} \ge l$}) }$.
		
		\Statex \textbf{Step 8:} $\textit{$I_{l}$} \gets$ If $p_{l} \le \alpha$, then $\forall o_{j}: o_{j} < o_{l}$ forms a multi-mode segment. We recurse into this interval and return all found  modal intervals. Else return $\phi$ i.e., an empty set. 
		% right
		\Statex \textbf{Step 9:} $\textit{$I_{r}$} \gets$ If $p_{u} \le \alpha$, then $\forall o_{j}: o_{j} > o_{u}$ forms a multi-mode segment. We recurse into this interval and return all found  modal intervals. Else return $\phi$ i.e., an empty set. 
	\Statex \textbf{Step 10:} The final set of all modal interval is $I_{l} \bigcup I_{mid} \bigcup I_{r}$.
		\Statex \textbf{Step 11:} As we knew that combo-SAD features have high positive value for speech and low value for different noises, the cluster with highest average feature value is taken as speech and rest clusters as non-speech. In some instances, where two prominent noise sources were present such as non-stationary background noise and occasional tonal impulsive noise, this approach led to three or more clusters.
%
%\EndFunction
%	\EndProcedure
	\end{algorithmic}
\end{algorithm}
%\vspce{5mm}
\setlength{\textfloatsep}{10pt}% Remove \textfloatsep	
\vspace{-4mm}
\subsection{Dip-based clustering}
\vspace{-2mm}
We used the dip test recursively to locate the modal intervals that could contain speech or non-speech frames. We explain the proposed clustering approach by looking at Figure~\ref{fig_illus} and going through the Algorithm~\ref{dipsad}~\textbf{\textit{Dip-SAD}}. Figure~\ref{fig_illus} illustrates a simulated scenario showing five categories in the feature space. The top sub-figure shows the histogram of features, while the bottom one shows empirical cumulative distribution. Clearly, the region R3, R4 and R5 lie close to each other in the feature space. On applying the clustering approach described in Algorithm~\ref{dipsad}~\textit{Dip-SAD}, the first modal interval detected consisted of R3, R4 and R5 (Step 3 in~\textit{Dip-SAD}). On recursing again in this interval for each $o_{j}$ such that $ o_{L} \le o_{j} \le o_{U}$, we get all the three regions R3, R4 and R5 that forms $I_{mid}$ i.e., middle modal intervals (Step 4). Next, we recurse into the right and left side of the primary interval to find if other segments were present (Step 5). While recursing to the left and right, we included the nearest detected modes from respective left or right region, i.e., for left recursion, region R3 in included in the search region while for right recursion, region R5 is included in the search region (Step 6). Thus, upper limit ($u$) for left search is minimum among all detected upper limits, i.e., upper limit of region R3. On the other hand, lower limit ($l$) for right search is chosen as maximum among all lower limits in detected regions, i.e., lower limit of R5 (Step 6). This strategy ensures that the left and right searches will either have unimodal (means same region extended in that direction such as R5 here extends till the end of the right region) or have multi-modalities (means different modes in that direction such as R1 and R2 in left). This is done in Step 6 of the Algorithm~\ref{dipsad}~\textit{Dip-SAD}. After we have upper limit, $u$ and lower limit, $l$ for left and right searches respectively, we iterate using Algorithm~\ref{computedip}~\textit{computeDip} on both regions to get the corresponding p-values, $p_{l}$ and $p_{u}$ (Step 7). From the corresponding p-values of such recursions, we conclude unimodality if $p_{l} > \alpha$ and return empty set $\phi$. If $p_{l} \le \alpha$, we find the corresponding modal interval and add it to set $I_{l}$ that is set of modal intervals for left region (Step 8). Similarly, we do for right search (Step 9) to get set $I_{r}$ that is set of modal intervals for the right region. The final set of all modal intervals is the union of middle set $I_{mid}$, left set $I_{l}$ and right set $I_{r}$. The Figure~\ref{fig_illus} was for illustration of the dip-based clustering approach. For speech activity detection (SAD), at the end of recursive dip tests on detected primary modal interval, left region and right region, we usually get just two, three or four clusters. We found that when there are more than one type of noise in an utterance such as non-stationary background noise, occasional impulsive noise~\emph{etc} then each non-speech region with a specific noise-type got clustered separately. 
%Such a scenario can be referred as having "multi-layer noise" as different noise sources appear as separate regions in feature space. 
From~\cite{sadjadi2013unsupervised}, we know that the \textit{Combo} features are relatively large positive values for speech and very small positive or negative values for noise. We leverage this fact in assigning clusters to speech or non-speech class. The cluster with highest average sample value was assigned to speech and rest clusters corresponded to non-speech. This assignment was done automatically on the basis of average feature value for each detected cluster. Authors in~\cite{graciarena2016sri} also noticed that the \textit{Combo} features for OpenSAD data were significantly tri-modal on some channels and tri-modal GMM helped in gaining improvements in DCF (Section 6.4.1)~\cite{graciarena2016sri}.
%%
%\skip\footins=\bigskipamount
\begin{table*}[!t]
\centering
\caption{DCF (\%) with two-second collar on all channels of Levantine Arabic (alv) in training set of NIST OpenSAD-2015 data. The scores were averaged over all audio files.}
\begin{tabular}{|c|c|c|c|c|c|c|c|}
\hline
\textbf{System} &$\mathbf{alv-B}$&$\mathbf{alv-D}$&$\mathbf{alv-E}$&$\mathbf{alv-F}$& $\mathbf{alv-G}$& $\mathbf{alv-H}$ & $\mathbf{alv-src}$ \\ \hline
%Baseline  & 8.54 & 7.21& 6.09 & 5.60&  1.51 & 6.07  & 3.02\\ \hline
Combo-SAD & 8.54 & 7.21 & 6.09& 5.60& 1.51& 6.07 & 3.02\\ \hline
Proposed & 13.21 & 6.40 & 5.83 &4.19 & 1.34  & 3.63 & 2.68    \\ \hline
%Proposed DipSAD & 2.11 & 6.25 & 1.64  & 5.16 & 1.45  & 4.01 & 2.68    \\ \hline
% 13.21 6.40 5.83 4.19 1.34 3.63 2.68
Relative Improvement (\%) & -54.68 & 11.23 & 4.27 & 25.18 & 11.26 & 40.20 & 11.26 \\\hline
%%   -54.6838   11.2344    4.2693   25.1786   11.2583   40.1977   11.2583
%
%Relative Improvement (\%) & 75.29 & 13.31 & 73.07 & 7.86& 3.97&33.94& 11.26 \\\hline	
\end{tabular}
\label{table_results_alv}
\end{table*}
\begin{table*}[!t]
\centering
\caption{DCF (\%) with two-second collar on all channels of American English (eng) in training set of NIST OpenSAD-2015 data.}
\begin{tabular}{|c|c|c|c|c|c|c|c|}
 \hline
\textbf{System} &$\mathbf{eng-B}$&$\mathbf{eng-D}$&$\mathbf{eng-E}$&$\mathbf{eng-F}$& $\mathbf{eng-G}$& $\mathbf{eng-H}$ & $\mathbf{eng-src}$ \\ \hline
%Combo-SAD & 9.65 & 10.32 & 6.44  & 5.83 & 2.36 & 5.66 & 4.18 \\ \hline
Combo-SAD & 9.65 & 10.32 & 6.44  & 5.83 & 8.18 & 5.66 & 4.18 \\ \hline
%Proposed DipSAD & 6.84 & 8.82& 2.65  & 4.84 & 2.32   & 4.14 & 4.11    \\ \hline
Proposed & 10.68 & 8.18 &5.12 & 2.96 & 9.30 & 4.11 & 6.87 \\ \hline
% 10.68 8.18 5.12 2.96 9.30 4.11 6.87
Relative Improvement (\%) &-10.67 & 20.74  &20.50 &49.23 & -13.69 &27.38 & -64.35 \\ \hline	
%Relative Improvement (\%) &-10.67 & 20.74  &20.50 &49.23 &-294.07 &27.38 & -64.35 \\ \hline	
%%-10.6736   20.7364   20.4969   49.2281 -294.0678   27.3852  -64.3541
%
%Relative Improvement (\%) & 29.12 & 14.53 &58.85 &16.98&1.69& 26.86 & 1.67    \\ \hline	
\end{tabular}
\label{table_results_eng}
\end{table*}
\begin{table*}[!t]
\centering
\caption{DCF (\%) with two-second collar on all channels of Urdu (urd) in training set of NIST OpenSAD 2015 data.}
\begin{tabular}{|c|c|c|c|c|c|c|c|}
\hline
\textbf{System} &$\mathbf{urd-B}$&$\mathbf{urd-D}$&$\mathbf{urd-E}$&$\mathbf{urd-F}$& $\mathbf{urd-G}$& $\mathbf{urd-H}$ & $\mathbf{urd-src}$ \\ \hline
Combo-SAD & 7.63 & 6.98& 5.69 & 5.76 & 3.73   & 5.62 & 3.48 \\ \hline
Proposed & 5.85 & 5.51 & 5.30 & 5.26 & 3.67 & 4.78 & 4.22   \\ \hline
% 5.85 5.51 5.30 5.26 3.67 4.78 4.22
%Proposed DipSAD & 5.83& 5.48& 5.12 & 5.24 &3.64  &4.75 & 3.06   \\ \hline
Relative Improvement (\%)& 23.33 & 21.06 &6.85 &8.68 &1.61 & 14.95 & -21.26\\ \hline	
%%
%   23.3290   21.0602    6.8541    8.6806    1.6086   14.9466  -21.2644
%Relative Improvement (\%) &23.59& 21.49& 10.02&9.03&2.41&15.48&12.07 \\ \hline	
%
\end{tabular}
\label{table_results_urd}
\end{table*}
\vspace{-3mm}
\section{Results \& discussions}
\vspace{-3mm}
We used 40ms windows with a 10ms skip-rate for extracting the \textit{Combo} features from each utterance. The sampling rate for processing the speech data was kept at 8kHz. The~\textit{NIST OpenSAD 2015} program was organized to advance the state-of-the-art SAD over extremely degraded communication channels~\cite{opensad2}. Six channels namely B, D, E, F, G and H from the DARPA RATS were included in the training set along with the source channel. This data consisted of re-transmitted telephone conversations captured through different communication channels. This data was provided at 16 kHz sampling rate with 16 bit resolution. We downsample the OpenSAD data to 8 kHz for feature extraction and further processing. In this study, we evaluate all channels of the training set as techniques being evaluated are fully unsupervised and parameter-free. 

Recently, NIST organized speech analytic technologies evaluation \textit{NIST OpenSAT 2017}~\cite{opensat}. It had three tasks: SAD, key word search, and automatic speech recognition. We evaluated the proposed SAD approach for OpenSAT public safety communications (PSC) data. It contained audio recordings from sofa super store fire (SSSF) dispatcher that occurred on June 18, 2007 in Charleston, South Carolina. The data constitute real fire-response operational data that can not be duplicated through controlled scientific collection~\cite{opensat}. Thus, the data is rich in naturalistic distortions such as (i) land mobile radio transmission effects; (ii) speech under cognitive and physical stress; (iii) varying background noise types and levels etc~\cite{opensat}. The data consisted of six audio recordings, each of approximately five minute duration, thus making up a total of 30 minutes of \verb!dev! data. The data were provided as 16-bit signed integer PCM at 8 kHz sampling rate. The \verb!dev! set was shipped with the ground-truth SAD reference labels for evaluation. %Using the ground-truth SAD annotations, we found that SSSF audio data contained overall 41.85 \% speech (by duration).

The evaluation metric used in NIST OpenSAD-2015 and NIST OpenSAT-2017 was the detection cost function (DCF) given by:
\begin{equation}
DCF = 0.25*P_{fa} + 0.75*P_{miss}
\end{equation}
where $P_{fa}$ is the false alarm rate (non-speech frames detected as speech) and $P_{miss}$ is the miss rate (speech frames detected as non-speech). The DCF values were computed for each audio file and averaged to get the DCF for each channel over three languages in NIST OpenSAD. We incorporated the two-second collar around each speech region in accordance with the NIST OpenSAD 2015 protocol. Table~\ref{table_results_alv}, Table~\ref{table_results_eng} and Table~\ref{table_results_urd} shows the comparison of results obtained with the proposed technique using a significance level, $\alpha=0.05$ and Combo-SAD baseline. The baseline Combo-SAD approach had \textit{Combo} features considered for fitting a two-component GMM. We chose 0.5 weights for both speech and non-speech GMM during threshold selection in baseline~\cite{sadjadi2013unsupervised}. Fixing the weights made the approach parameter-free. Clearly, we can see that the proposed approach led to significant relative gains in DCF as compared to the baseline Combo-SAD except for alv-B, eng-B, eng-G, eng-src, urd-src channels. The Combo-SAD baseline is a model-based technique and it performs well when Combo features are bi-modal. On channels where Combo-SAD is better than the proposed Dip-SAD approach, we found that Combo feature were distinctly bi-modal for majority of the utterances. Overall, we found that the Dip-SAD had reasonable DCF gains over Combo-SAD on many channels. The poor performance of Dip-SAD on some channels is possibly due to over-clustering of speech into two clusters. In future, we would consider semi-supervised cluster assignments for such cases. Table~\ref{table_opensat_results} shows the DCF with \textit{no collar} for all audio recordings in NIST OpenSAT PSC SSSF \verb!dev! set. We can see that the proposed Dip-SAD approach has overall 3.89\% relative improvement in DCF as compared to GMM baseline with same features.
%(Combo-SAD). 
% As number of audio files is different for three language on a given channel, we have different Combo-SAD statistics. 
%
%OpenSAT data has variety of distortions, noise and reverberation that made the SAD challenging. 
%GMM assume a bi-modal gaussian model in feature space that is clearly not the case for these six utterance. 
%
%Table~\ref{table_avg_results2} shows the channel-wise average gains in DCF. Here, the average is calculated over all audio files in each one of the three languages.
%The weight, $\gamma =0.25$ put more weight to miss rate
%
\begin{table}[!t]
\centering
\caption{DCF with no collar on all audio recording in PSC SSSF dev set from NIST OpenSAT 2017.}
\begin{tabular}{|c|c|c|c|}
\hline
Audio & GMM  & Proposed & Relative \\ 
%\hline
\text{name}& (\%)& (\%) & Improvement(\%) \\ \hline
\verb!sssf_dev_001! & 10.04 & 8.76 & 12.75 \\ \hline
\verb!sssf_dev_002! & 9.25 & 11.03 & -19.24 \\ \hline
\verb!sssf_dev_003! & 6.20 & 5.67 & 8.55\\ \hline
\verb!sssf_dev_004! & 4.39 & 4.57 & -4.10 \\ \hline
\verb!sssf_dev_005! & 6.58 & 5.13 & 22.04\\ \hline
\verb!sssf_dev_006! & 8.29 & 7.88 & 4.95 \\ \hline
Overall & 7.46 & 7.17 & 3.89 \\ \hline
\end{tabular}
\label{table_opensat_results}
\end{table}
%
% Different frequency bands (HF, UHF and VHF) with modulation types (narrow-band FM, wide-band FM, AM, frequency-hopping spread-spectrum and SSB) from the DARPA RATS data were re-organized to form the NIST OpenSAD 2015 data (see Table~\ref{table_nist_channels}). 
%two additional channels (A and C) in the evaluation set.On this set, the average duration of an audio recording is 15 minutes and 19 seconds, with speech regions comprising 35.12\% of the audio.
%
%The NIST 2015 OpenSAD evaluation uses data collected for the DARPA RATS program. The organizers divided their data
%into three parts, Training, Development and Evaluation, all
%
%(see Table 2). The
%source audio originated from existing LDC corpora (Fisher English, Fisher  and CALLFRIEND Farsi) and data collected in the RATS program.
%
%Training and Development parts include speech in five languages: Levantine Arabic
%(alv), American English (eng), Farsi (fas), Pashto (pus) and Urdu (urd). Channels A and C were excluded from the Training and Development parts by the organizers.
%
%The Development set provided by NIST was divided into two parts, dev-1 and dev-2, the latter having additional channels. Since the ground-truth annotations were deemed reliable by NIST only for channels B, D, E, F, G and H in dev-2, we excluded channels XA through XN
%
\vspace{-4mm}
\section{Conclusions}
\vspace{-2mm}
This study leverages Hartigan dip test for unsupervised speech activity detection for scenarios that lack  annotations. We used Combo features in proposed clustering approach as these were found to perform well on extremely noisy DARPA RATS data. The proposed approach is deterministic and parameter-free. Results on NIST OpenSAD-2015 data shows proposed approach to be significantly better than the baseline on many channels from three languages. The overall relative improvement in DCF was 3.89\% for NIST OpenSAT. 
%Recursions based on dip test were used for locating the prominent modes in the feature space.
%
%Future work would focus on extending this approach for other speech tasks such as domain adaptation.
%\vspace{-4mm}
%
 %]] an extended set of speech features for speech activity detection using the proposed approach. 
%\section{Acknowledgements}
%%
%The ISCA Board
%
%\newpage
%\bibliographystyle{IEEEtran}
\bibliographystyle{plain}

%%\nocite{*}
%\bibliography{dipsad}
\bibliography{icassp}
\end{document}